\documentclass{elsarticle}

\usepackage{graphicx} 
\usepackage[colorlinks=true,bookmarks=true,bookmarksopen=true,breaklinks=true]{hyperref}

\begin{document}

\begin{frontmatter}

\title{Combined surface acoustic wave and surface plasmon resonance 
measurement of collagen and fibrinogen layer physical properties}

\author{J.-M. Friedt}
\address{FEMTO-ST Time \& Frequency, and SENSeOR SAS, \\Besan\c con, France ({\tt jmfriedt@femto-st.fr})}
\author{L.A. Francis}
\address{Sensors, Microsystems and Actuators Laboratory of Louvain (SMALL),
ICTEAM Institute, \\
Universit\'e catholique de Louvain (UCL), Belgium}

\begin{abstract}
We use an instrument combining optical (surface plasmon resonance) and 
acoustic (Love mode surface acoustic wave device) real-time measurements on a same 
surface for the identification of water content in collagen and fibrinogen
protein layers. After calibration of the surface acoustic wave device 
sensitivity by copper electrodeposition and surfactant adsorption,  
the bound mass and its physical properties -- density and optical index -- 
are extracted from the complementary measurement techniques and lead to 
thickness and water ratio values compatible with the observed signal 
shifts. Such results are especially usefully for protein layers with a 
high water content as shown here for collagen on an hydrophobic surface. 
We obtain the following results: collagen layers include 70$\pm$20\% water 
and are 16$\pm$3 to 19$\pm$3~nm thick for bulk concentrations ranging from 
30 to 300~$\mu$g/ml. Fibrinogen layers include 50$\pm$10\% water for layer 
thicknesses in the 6$\pm$1.5 to 13$\pm$2~nm range when the bulk concentration 
is in the 46 to 460~$\mu$g/ml range.
\end{abstract}


\end{frontmatter}


\section{Introduction}

Sorption processes at the solid/liquid interface by which (bio)molecules
bind to material surfaces are of interest for biosensors, biomaterials, 
material and surface science. Understanding the three-dimensional 
organization (including density, solvent content and thickness) of the 
resulting sorbed film and its evolution during the adsorption process is crucial 
for many applications in these domains. For biosensors, more specifically, 
there is a need to monitor the response in real-time, in order to assess
the adsorption kinetics, and to be able to 
distinguish contributions coming from the dry sorbed mass, which is the 
physical criterion for estimating sensitivity, and those that should be
attributed to effects intimately associated to the layer organization, like 
sorbent-bound water and hydrodynamic effects for example. While a wide variety of 
methods can qualitatively detect the formation of the sorbed film, almost 
none of them, taken alone, is quantitative and able to reveal the film organization, 
and only few can monitor the process in real-time. Scanning probe microscopies 
might fulfill all these requirements, mainly for submonolayers sorbed on 
smooth surfaces \cite{book2}. Neutron reflectivity \cite{book1}, X-ray Photoelectron
Spectroscopy (XPS) \cite{XPS}, mass spectroscopy and radiolabeling are 
quantitative techniques able to directly measure the dry sorbed amount.
Of all the direct detection ({\it i.e.} label-less) techniques, 
we have identified acoustic and optical methods as being the only ones 
fulfilling two fundamental criteria of our measurements: time resolved and 
in-situ (liquid phase) measurement of the physical properties of the adsorbed 
layer.

Various methods of direct detection of
biochemical layers have been developed, either based on the disturbance
of an acoustic wave \cite{thompson} 
(quartz crystal microbalance \cite{gizeli3} -- QCM -- and 
surface acoustic wave devices \cite{gizeli2} -- SAW) or of an evanescent
electromagnetic wave (optical waveguide sensors \cite{waveguide1,waveguide2}, 
surface plasmon resonance \cite{liedberg,gizeli1} -- SPR). 
While each one of these transducers individually provides reliable qualitative curves 
during protein adsorption on their functionalized surfaces, extraction of 
quantitative physical parameters such as optical index, density, viscosity 
or water content requires modeling of the adsorbed layers \cite{kasemo3}. 
The modeling includes multiple parameters which must be identified 
simultaneously: hence the need for the combination of (acoustic and optical)
detection methods in a single instrument \cite{bailey,sspr,sawspr,sawspr2}
to separate contributions as a same layer is reacting with the surface under
investigation. Multiple investigators have identified such a combination
of measurement methods as fruitful means of extracting independent physical
properties of adsorbed layers, including the challenging combination of
QCM and SPR 
\cite{keller2000formation,bund2003combining,reimhult2004simultaneous,reimhult2006multitechnique,zong2008quartz} 
or comparing the results of successive experiments using different instruments 
\cite{larsson2003characterization,malmstrom2007viscoelastic,ansorena2011comparative,konradi2012using}, 
combining QCM and reflectometry \cite{edvardsson2008qcm} or measuring 
separately using the two techniques \cite{richter2005following}, or SAW and SPR
\cite{bender2009development}.

We here use a combination of Love mode SAW device and SPR to identify values 
of density, water content and thickness of surfactant films and protein layers
(collagen and fibrinogen) adsorbed on methyl-terminated surfaces.
This combined measurement is necessary when attempting to convert a raw signal
as observed at the output of a transducer (angle shift for SPR, phase
and magnitude shift for SAW or frequency and damping for a QCM) to the actual 
protein mass bound to the surface, which
is the physical criterion for estimating the expected highest possible
sensitivity of a biosensor since it provides an estimate of the density of
active sites on the surface.
We furthermore compare the signals obtained from quartz crystal microbalance 
with dissipation monitoring
(QCM-D \cite{QCMD2,QCMD1}) measurements to that of the SAW
and, based on the results obtained from the analysis of the SAW/SPR combination, show how 
SAW and QCM interact differently with the layer. The QCM displays a strong 
sensitivity to viscous interactions with adsorbed layers as was shown 
previously \cite{voinova,kasemo2}. SAW devices are sensitive
to mass loading, visco-elastic interactions and electrical charge
accumulation on the sensing area \cite{ricco}, but with different influences
due to the different frequencies and hence penetration ratio of the shear
acoustic wave with respect to the layer thickness. 

Love mode surface acoustic waves were chosen for their high mass sensitivity and
their compatibility with measurements in liquid media
\cite{mchale,gizeliigg,wang,du}. Being based on the
propagation of a shear horizontal acoustic wave, their interaction with
the surrounding liquid is restricted to an evanescent coupling with the 
viscous liquid. Although bulk liquid viscosity properties 
affect the acoustic wave propagation \cite{viscojakoby,herrmann}, 
including the phase shift of Love mode SAW \cite{lamia}, we will
throughout this investigation consider that the phase shift affecting
the SAW device is solely related to adsorbed mass and not to viscous
effects of the adsorbed layer, in order to reduce the number of 
unknowns. Such a crude assumption could be removed by exploiting the
SAW insertion losses, 
for introducing the adsorbed layer viscosity. Throughout the analysis
proposed here, we consider hydrodynamic interactions of the acoustic wave
with the solvent filled adsorbed layer, as well as the equivalent
optical index of the protein-solvent mixture, without focusing on the
dynamic viscosity of this adsorbed layer but only on the viscosity of the 
fluid yielding shear wave evanscent coupling with the solvent.

The chosen protein layers consist of collagen and fibrinogen, selected as
references both for their interest in engineering biocompatible
surfaces \cite{prothese,mara} but most significantly for their strong solvent
content and hence acoustic properties challenging to analyze.
Collagen is a fibrillar protein of the extracellular matrix possessing
self-assembly properties, involved in biorecognition processes. The
collagen macromolecule consists of a triple helix with dimensions about 
300~nm in length and about 1.5~nm in diameter, and weights about 300 
kilodaltons \cite{colla}. 
Organization of collagen films adsorbed on hydrophobic
surfaces - CH$_3$ -- terminated self-assembled monolayers (SAMs) and polystyrene 
-- from 30 to 40~$\mu$g/ml -- were fully characterized under
water or after drying using AFM \cite{denis,cupere}, X-ray photoelectron 
spectroscopy \cite{dufrene} and radioassays. By AFM scratching experiments, thicknesses of
the film adsorbed on methyl-terminated surfaces was estimated to be about
20~nm under water and 7-8 nm after drying. Furthermore, it was found that
the measurement was strongly influenced at weak applied forces ($<$0.5 nN),
in relation with the long-range repulsion ($\sim$50 to $\sim$250 nm) observed by
AFM force-distance curves, strongly suggesting that at least some molecules
of the film must protrude into the solution \cite{cupere}. Adsorbed
amount of (dry) collagen on methyl-terminated surfaces were estimated
to be between 0.4 and 0.8 $\mu$g/cm$^2$ by combining AFM and XPS measurements.
Values near 0.5~$\mu$g/cm$^2$ were determined to be adsorbed on polystyrene by
radiolabeling. 

Fibrinogen, on the other hand, is a blood protein that presents
three globular domains linked together by fibrilar segments. Similarly to 
collagen, its molecular weight is about 340~kilodaltons.
Fibrinogen adsorbed on various surfaces has been imaged by 
AFM as well \cite{drake}, down to molecular resolution \cite{fib0,fib1,
fib2,fib3}. 

\begin{figure}[h!tb]
\includegraphics[width=\linewidth]{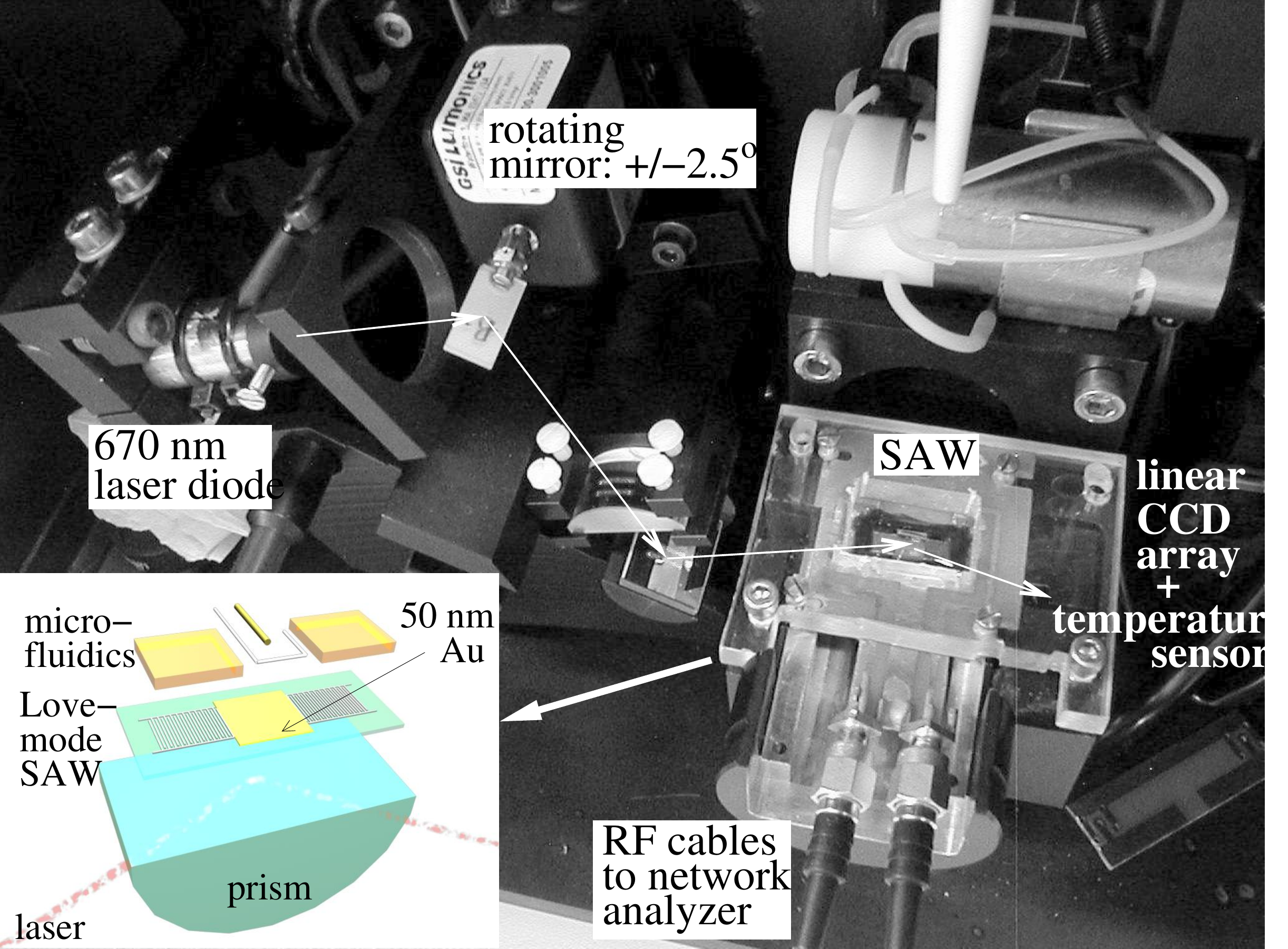}
\caption{Picture of the experimental setup combinging a SPR instrument
(670~nm-wavelength laser with mechanical sweep of the illumination angle)
with a piezoelectric (ST-cut quartz) substrate acting as both SPR sensing 
area since coated with a 50-nm~Au layer and propagating a Love-mode SAW.
Bottom-left inset: schematic view of he measurement head, emphasizing the need
for integrated microfluidics to prevent liquid from reaching the interdigitated
transducers. In this schematic chart, an electrochemical application is considered
by showing the (bent) counter electrode and (straight) reference electrode located
over the gold-coated sensing area acting as working electrode.}
\label{toto}
\end{figure}

In this presentation, similar films will be investigated, but under the
new perspective of combined acoustic and optical measurements for 
extracting the thickness, mass, and solvent density in a buffer medium.
While the SAW/SPR technique (Fig. \ref{toto}) and data processing with 
our proposed formalism
was already shown \cite{analchem}
to be appropriate in the identification of some physical 
properties of a rigid adsorbed layer (S-layer) \cite{JAP}, we here 
quantitatively analyze the organization of collagen and fibrinogen adsorbed 
layers which are expected to possess a substantial water content.
This condition is expected to lead to the largest differences between
acoustic and optical signals since both techniques respond differently
to a viscous (solvent containing) layer: acoustic methods tend to
overestimate the bound mass due to hydrodynamic interactions, while optical
methods provide an estimate of the dry bound mass after appropriate 
modeling of the response but cannot resolve both parameters, thickness
and optical index of the layer which SPR is sensitive to. This optical
index can be assumed to be the weighted value of the index of water and that
of proteins to the volumic part that these components occupy in the film.
SPR response is thus dependent on the film organization and on the dry
adsorbed amount. 

Unlike the model used in other studies \cite{viscous_gizeli}, we
here assume that the SAW device signal shift is predominantly due
to added rigidly bound mass on the electrode.  Indeed we have
shown that, while the QCM is sensitive via hydrodynamic interactions with the
topography induced by rough copper electrodeposition \cite{ECS}, SAW is much
less sensitive \cite{me}. The contribution of the wave coupling with
the viscous fluid to the phase shift \cite{lamia} 
will be neglected throughout this investigation.
Indeed, collagen films were analyzed using SAW/SPR after different conditionings, 
inducing a different film organization and related viscoelastic 
behavior (as probed by AFM and QCM respectively) for a similar adsorbed 
amount. These differences in the film properties were found to result in only 
minor changes in the SAW response. For all these reasons, the SAW phase 
response is considered in this
article to be only mass dependent during adsorption phenomena. 
Hence, we use 
a proportionality relationship between the mass of the layer $\Delta m$ per unit area $A=5\times 5.5$~mm$^2$ here
($\Delta m/A$) -- including the rigidly bound water -- and the frequency 
shift: $\frac{\Delta m}{A}=\frac{\Delta f}{S\times f_0}$ where $f_0$ is the
frequency at which the phase is monitored in an open-loop configuration, 
$\Delta f$ is the frequency shift
obtained after conversion from phase to frequency shift through the
experimentally measured phase to frequency linear relationship, and
$S$ is the mass sensitivity calibrated by copper electrodeposition. Since
the mass sensitivity calibration is a central part in extracting quantitative 
results from the experimental data, we confirm the results obtained with copper
electrodeposition by measuring the SAW signal change during adsorption
of a surfactant, cetyltrimethylammonium bromide (CTAB). 
Hence, four sets of experiments will be described in the experimental
section below, after presenting the instrument setup:
\begin{enumerate}
\item experimental calibration of the gravimetric sensitivity of the SAW
transducer in liquid phase is a core step to the quantitative analysis: such
a step is performed by reversible copper electrodeposition, allowing for
multiple cycles under different deposited mass configurations to be repeated,
\item because the mechanical properties and surface roughness of copper differ
from the biofilm we are investigated, the calibration is validated using CTAB
monolayer deposition,
\item collagen thick films will be assembled on the surface under different bulk
concentrations,
\item fibrinogen thick films will be assembled on the surface under different bulk
concentrations.
\end{enumerate}

\section{Experimental section}

{\bf Instrumentation} 
The combined SAW/SPR instrument developed for performing this experiment has 
been described previously \cite{JAP}: a modified Ibis II SPR instrument
(IBIS Technologies BV, Netherlands) is used to inject a 670~nm
laser in a quartz substrate (Kretschmann configuration) and monitor
the reflected intensity {\it vs.} angle with an accuracy of
$\pm 2.555^\circ$/200~pixels at two locations separated by about 2~mm
on the sensing surface. The ST-cut quartz substrate is patterned 
with double-fingers interdigitated electrode for launching a
Love mode acoustic wave at a center frequency of 123.5~MHz. The guiding
layer is made of a 1.13~$\mu$m thick PECVD silicon dioxide layer.
The phase and insertion loss of the acoustic wave device are
monitored using an HP~4396A network analyzer while a custom software
records the full reflected intensity {\it vs.} angle SPR curve.
The $5\times 5.5$~mm$^2$ sensing area is coated by e-gun evaporation
with nominal thicknesses
of 10~nm $Ti$ and 50$\pm$5~nm $Au$ to support surface plasmon resonance,
and is either grounded to minimize varying salt concentration effects 
on the signal measured by the SAW device during biological experiments,
or connected as the working electrode during copper electrodeposition
calibration through a bias-T circuit. Both sets of data -- SAW phase at a given frequency and
SPR reflected intensity curves -- are time stamped with an accuracy of
1 second on the common personnal computer internal 
clock reference for future processing, which for the SPR curve involves
polynomial fitting for improving 100-fold the accuracy on the localization
of the reflected intensity minimum. The temperature coefficient of the Love mode device
has been measured to be 34$\pm1$~ppm/K, in agreement with values reported in 
the literature \cite{du}, and the insertion loss is in the 
-22~dB to -30~dB range when the grounded sensing area is covered by a liquid. 
Both dual-finger interdigitated transducers (wavelength: 40~$\mu$m) are
protected from the liquid by a 100~$\mu$m-high SU8 wall capped with a piece
of quartz forming sealed protective chambers \cite{pack}. 
The sides of the SAW devices
are glued to an 7$\times$17~mm opening in a 1.6~mm-thick epoxy (FR4) printed 
circuit board (PCB) with the bottom area of the SAW tangent to the PCB,
allowing optical contact between the bottom surface of the SAW device and
a cylindrical prism through optical index matching oil ($n_{oil}=1.518$).
After wire bonding the SAW device to the PCB, the region surrounding the SAW
device is coated with 1~mm wide, 1-2~mm thick epoxy walls using H54
two-parts epoxy glue (Epoxy Technology, Billerica, MA, USA), forming a 180 
to 200~$\mu$l open well in which the solution is injected using a micropipette.

\begin{figure}[h!tb]
\includegraphics[width=\linewidth]{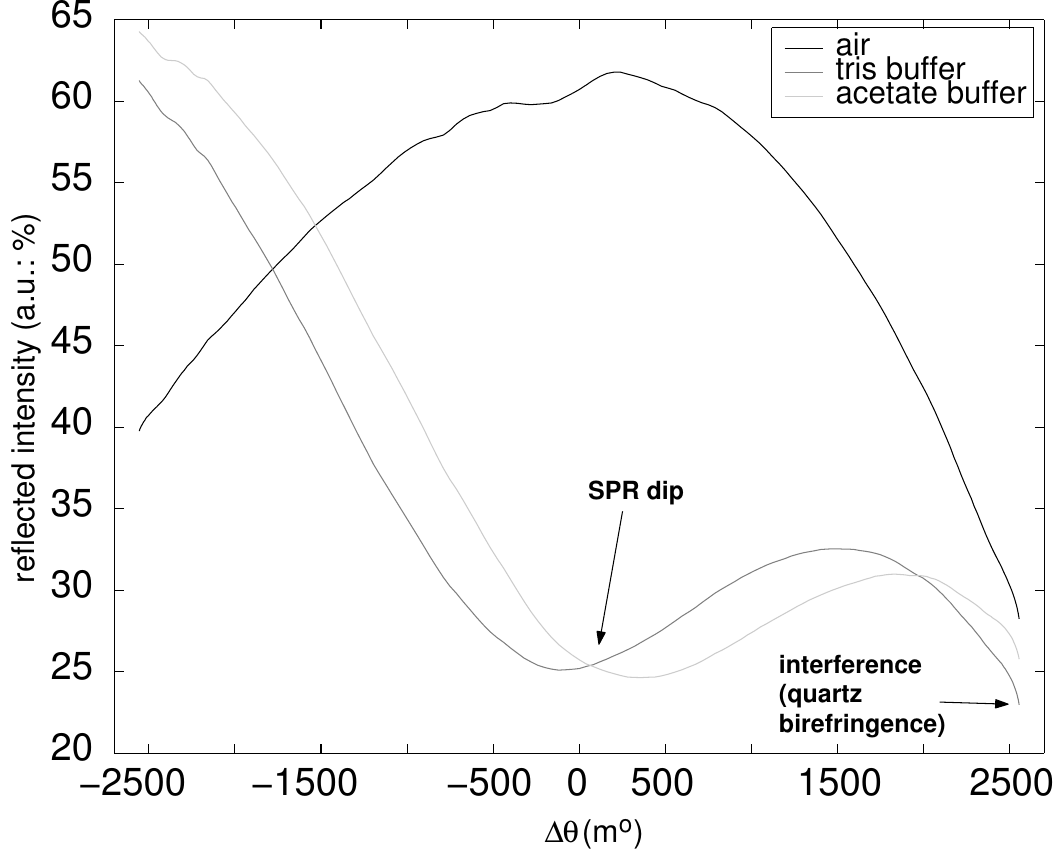}
\caption{
SPR dip observed on a quartz substrate coated with 
1.13~$\mu$m~$SiO_2$ and 10+50~nm $Ti+Au$. Notice the additional features on 
both angle extrema of the curve attributed to interference from quartz 
birefringence.}
\label{figdip}
\end{figure}

While both SPR and SAW sensors
display theoretically similar sensitivities and detection limits \cite{lod},
the acoustic 
sensor is extremely sensitive to temperature variations and care must be 
taken to properly control the environment around the instrument to keep 
the temperature stable \cite{bailey}.

Recording all SPR curves of the reflected intensity {\it vs.} angle
during the experiment for post-processing is required since the shape
of the dip is unusual (Fig. \ref{figdip}) and cannot be processed properly 
by the commercial software provided with the instrument which expects the 
substrate to be glass. As seen in Fig. \ref{figdip} in which the
reflected intensity {\em vs.} angle curve is displayed, the periodic dips 
observed when the gold coated SAW sensor is placed over the half-cylinder 
prism {\em in air} is attributed to the birefringence of quartz which is the
cause of the interference pattern after reflection of the laser on the metallic 
surface \cite{JAP}. The orientation of
the quartz substrate on which the SAW propagates has been chosen on purpose
to separate as much as possible the successive interference minima. The dip 
due to SPR is identified when the surface is coated with water, and the 
position of this dip only is monitored over time. A typical raw SPR reflected 
intensity {\em vs.} angle curve is displayed in Fig. \ref{figdip}: the mean
incidence angle is about 72$\pm$2$^\circ$ (not calibrated in the Ibis II
instrument) and the abscissa is graduated in relative angle shift.

The QCM-D measurements were performed using the electronics supplied
by Q-Sense AB (Sweden) using AT-quartz plano-plano resonators (Chintele Quartz
Technology Co. Ltd, Zhejiang, P.R.  China) coated, by evaporation, with 10~nm $Ti$ and 
50~nm $Au$, key-hole shaped counter electrode and fully coated sensing
electrode,  leading to a fundamental resonance frequency around 4.7~MHz 
and a quality factor in liquid around 5500. These experiments were performed
both in a lab-made open cuvette and in the commercially supplied flow cell
(with static liquid during the measurement). Both sets of measurements
are compatible. However, since the QCM setup is distinct from the SAW/SPR
setup, we will not attempt any kinetics comparison between both datasets.

{\bf Chemicals and Materials}
The $Cu$ electrodeposition step was performed using a solution of
$10^{-2}$~M $CuSO_4$ in $10^{-2}$~M $H_2SO_4$. The counter electrode
was a 0.25-mm-diameter platinum wire shaped in 3/4 of a circle and the
pseudo-reference electrode was made of a 1~mm-diameter 99.98+\% 
copper wire (Huntingdon, UK), cleaned by briefly dipping in 70\% nitric
acid before use. The electrochemical reaction was controlled by a PC3-300 
Gamry Instruments (Warminster, USA) potentiostat. 

Collagen was supplied by
Roche (Boehringer-Mannheim, Mannheim, Germany) as a sterile acidic solution
of 3~mg/ml solution of collagen type I from calf skin. It was diluted
in PBS buffer (137~mM $NaCl$, 6.44~mM $KH_2PO_4$, 2.7 mM $KCl$ and 
8~mM $Na_2HPO_4$) to reach the wanted concentration just prior to use. 
Type I fibrinogen from human plasma was obtained from Sigma-Aldrich. 
Both QCM and SAW/SPR $Au$ sensing electrodes were coated with an 
octadecanethiol hydrophobic self-assembled layer, 
leading to a contact angle above $110^\circ$. Alkanethiols were self-assembled
by dipping the QCMs in a 10$^{-3}$~M octadecanethiol (Sigma-Aldrich) solution 
in ethanol after a
15~minutes cleaning step in UV-O$_3$ and rinsing with ethanol. The SAW
devices were placed in a chamber with an ethanol saturated atmosphere, and
150 to 180~$\mu$l of the same thiol solution was deposited on the open well
above the sensing area. Adsorption was subsequently allowed for 3~hours,
after which the surfaces were rinsed with ethanol and dried under a stream of
nitrogen. The thiol solution and later biochemical
solutions are only in contact with the chemically inert H54 epoxy glue and
quartz slides capping the SAW device and never with the FR4 epoxy and 
copper of the PCB, thus minimizing risks of contamination of the solutions
and of the sensing surface.

\section{Results}

\subsection{SAW sensitivity calibration}

\begin{figure}[h!tb]
\includegraphics[width=\linewidth]{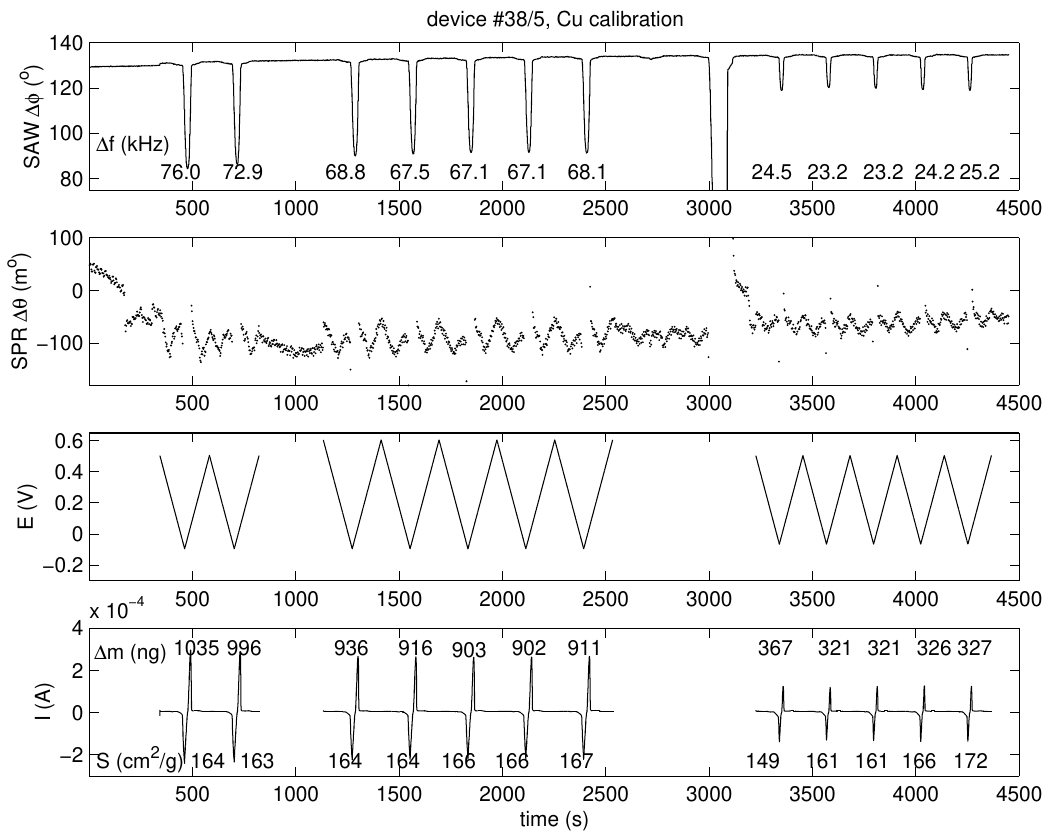} 
\caption{
Copper electrodeposition calibration curve. The resulting
experimentally measured sensitivity indicated on the bottom graph is
165~cm$^2$/g. From top to bottom: raw SAW phase shift and converted 
frequency shift via the phase to frequency linear relationship observed
on the Bode plot; SPR angle shift showing fluctuations with the potential
sweep due to electroreflectance effects; voltage applied to the potentiostat
({\it versus} a copper wire pseudo-reference electrode); current flowing
through the potentiostat and, on top, the mass deposited on the sensing
area obtained by integrating the current over time and converting to a
mass of deposited copper, and bottom the experimentally measured sensitivity 
calculated knowing the frequency shift (top graph) and the deposited mass
(bottom graph).
}
\label{Cu}
\end{figure}

Figure \ref{Cu} illustrates the calibration procedure using copper
electrodeposition: a cyclic voltametry sweep from +600~mV to -60 to -100~mV
(depending on the mass of copper to be deposited) {\em vs} a copper
wire acting as a pseudo-reference electrode provides an estimate of the
deposited mass, assuming a 100\% efficiency of the electrochemical process.
A current $I_j$ is monitored by the potentiostat at a sampling rate of 5~Hz, 
$j$ being a sample index. 

By selecting the voltage range for which the
current is negative (copper deposition), we obtain the deposited mass by
numerical integration
$$M_{Cu}=\frac{\Sigma_j I_j\times \delta t}{F}\times\frac{m_{Cu}}{n_e}$$ where
$F$ is the Faraday constant (96440~C), $\delta t=0.2$~s is the time
interval between two samples, $m_{Cu}=63.5$~g/mol is the molar weight of 
copper and $n_e=2$ the number of electrons required for copper reduction.
Simultaneously the phase shift $\Delta \phi$ of the SAW device due to the added
mass is measured and converted to a frequency variation $\Delta f$ via
the linear phase to frequency relationship measured on the Bode plot:
$\Delta f/\Delta \phi=1680\pm5$~Hz/$^\circ$ in our case.
Once the deposited mass is estimated, the SAW sensitivity $S$ is computed
considering the center oscillation frequency $f_0=123.2$~MHz, the working
electrode area $A$ and the frequency shift $\Delta f$ due to the added mass:
$S=\frac{\Delta f}{f_0}\times\frac{\Delta m}{A}$, leading to an experimentally 
measured sensitivity of 165$\pm$10~cm$^2$/g for one SAW device and 
145$\pm$15~cm$^2$/g for another similar sensor (data not shown), the 
difference being related to packaging issues and mostly the influence
of the protective SU8 wall on the acoustic path. 

Both values were confirmed with CTAB adsorption on octadecanethiol coated 
gold at a bulk concentration of 1.3$\times$10$^{-4}$~M: the surface density
of surfactant molecules upon adsorption on gold is known from the surface area per
molecule of 50~\AA$^2$ \cite{CTAB} and the molar weight of 364~g/mol, 
a mass per unit area of 120~ng/cm$^2$ is deduced assuming the formation of 
a monolayer. The monolayer formation lead to a frequency shift 
of the SAW device of 2.5~kHz from which a surface density of 125~ng/cm$^2$ is 
obtained. Interestingly, the simultaneously observed SPR angle shift can only
be interpreted with a layer of optical index at least 1.5, which
leads to a layer thickness of 1.5$\pm$0.5~nm fully coated with CTAB (no
solvent) after simulation using the formalism described in \cite{vigoureux}
and following the procedure described in more details later in this article.
The large uncertainty on the layer thickness is related to the low
molecular weight of CTAB leading to a small mass increase which leads to a
phase shift of only about 5 times the noise level of the SAW phase.
Such a high optical index is compatible with that provided in the literature 
for thiol monolayers as an extension of the optical index of bulk alkane 
solutions \cite{wim}. The surface mass density is compatible with that 
observed with QCM-D measurements following the assumption of a rigid mass, 
as validated by the good overlap of curves obtained at overtones 3, 5 and 7 
when normalized by the overtone number, and the low dissipation increase 
($\Delta D<1\times 10^{-6}$).

\subsection{SAW/SPR measurements}

\begin{figure}[h!tb]
\begin{tabular}{cc}
\includegraphics[width=0.49\linewidth]{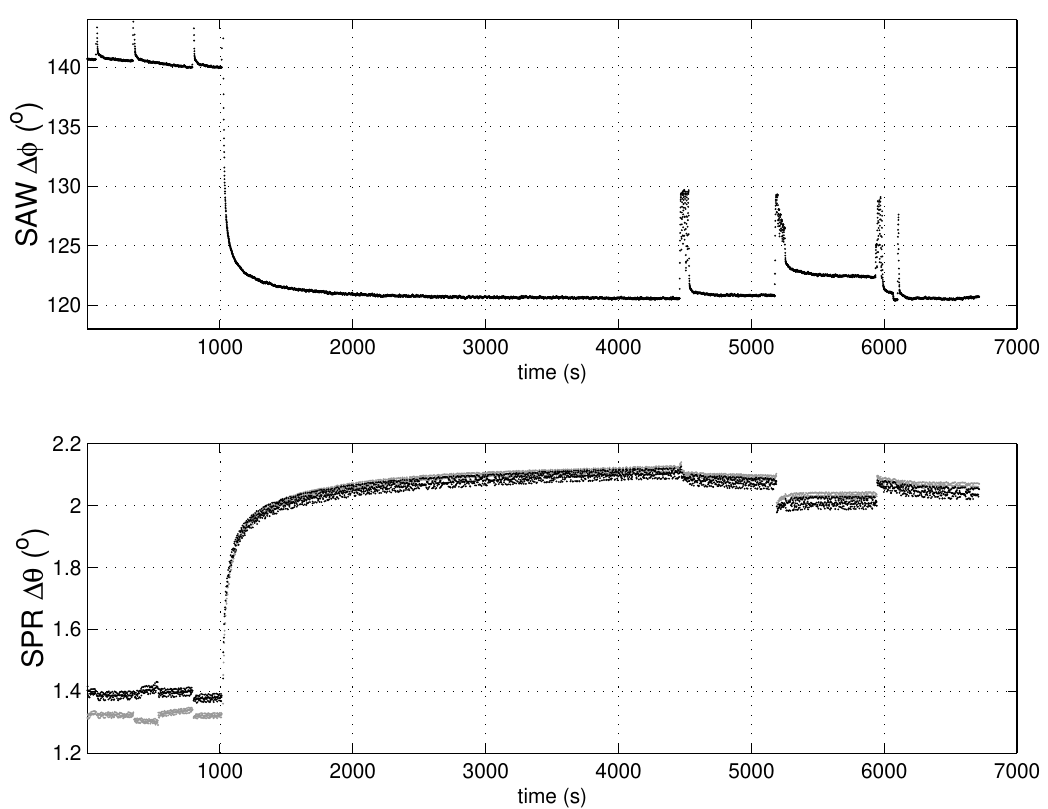} & 
\includegraphics[width=0.49\linewidth]{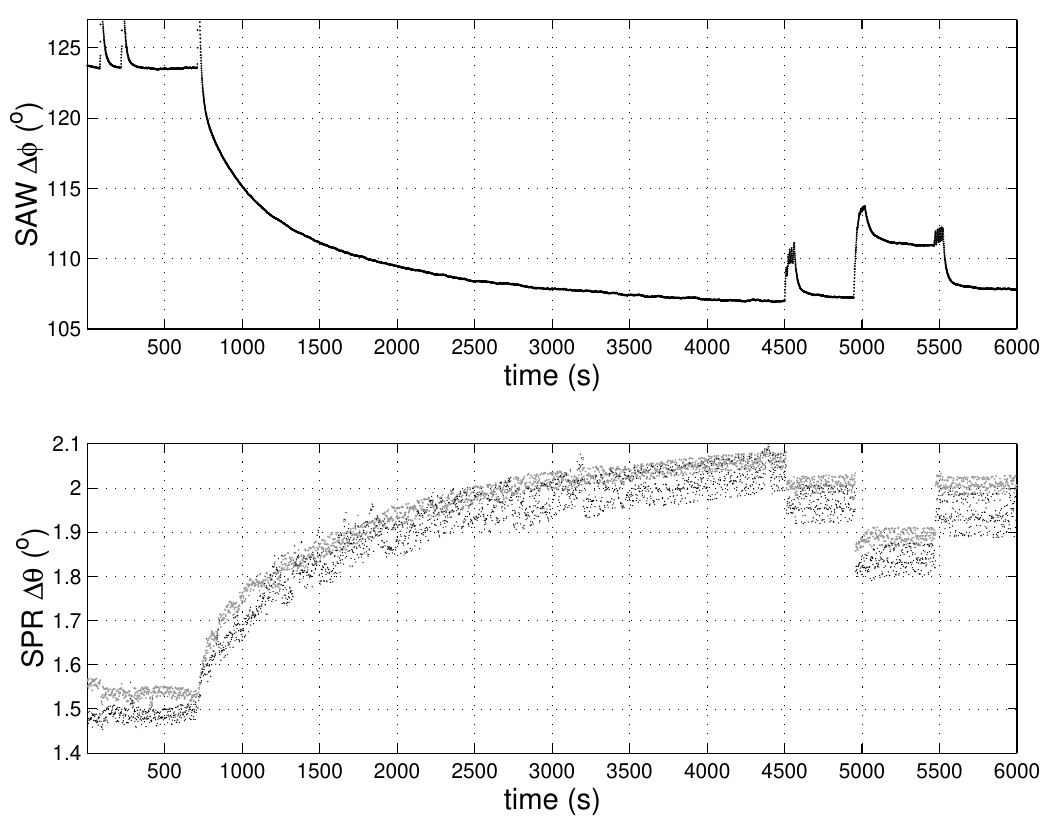} \\
(a) & (b) \\
\end{tabular}
\caption{(a) 300~$\mu$g/ml and (b) 30~$\mu$g/ml collagen adsorption curves. 
Top: phase shift monitored at the constant frequency of $f_0$=123.2~MHz, which can 
then be converted to an equivalent frequency shift as would be observed 
in a Phase-Locked Loop oscillator configuration via the linear phase to 
frequency relationship. 
During experiment (a), collagen adsorption lead to an insertion loss increase
of 6.5~dB, from 22.9~dB when the sensing area is covered with buffer to 29.4~dB
after collagen adsorption. During experiment (b), this insertion loss increased by
3.3~dB, from 24.4~dB prior to collagen adsorption to 27.7~dB after adsorption.
Bottom: SPR dip position angle shift. The sequence is PBS buffer (3 solution exchanges in
(a), 2 solution exchanges in (b)), collagen in the same PBS buffer, PBS 
rinsing step, DI water and PBS.}
\label{col}
\end{figure}

Once the mass sensitivity of the SAW device is calibrated, 
the combined SAW and SPR measurements are used to monitor protein
adsorption and extract the physical properties of the resulting films. 
Collagen and fibrinogen adsorption were performed on methyl-terminated
surfaces of the SAW/SPR device at concentrations of 30 or 300~$\mu$g/ml
and 46 or 460~$\mu$g/ml respectively. The sequence of solutions brought
in contact with the sensing surface was successively PBS (or PBS-water-PBS),
protein in PBS, PBS, water and finally PBS. Figure \ref{col} displays 
a typical signal observed during collagen adsorption on an hydrophobic 
surface, for (a) 300~$\mu$g/ml and (b) 30~$\mu$g/ml bulk concentrations.
Immediately after the conditioning with the collagen solution, both SAW and SPR
show a clear response, reaching after about one hour a clear plateau or a
beginning of a plateau, depending on the conditioning collagen
concentration. 

The SAW and SPR shifts at the end of the adsorption with a collagen 
concentration of 30~$\mu$g/ml were respectively 17.0$\pm$0.7~$^\circ$ and 
0.39$\pm 0.07$~$^\circ$, while a concentration of 300~$\mu$g/ml lead to 
19.2$\pm$0.5~$^\circ$ and 0.65$\pm 0.1$~$^\circ$ for SAW and SPR respectively. 
Adding PBS, after
collagen conditioning, resulted in very small loss ($<1$~$^\circ$ for SAW and
$<0.05$~$^\circ$ for SPR) of the response, stable with time, and subsequent
washing by water and PBS again provoked no signal change. This means that 
molecules do not desorb with time once adsorbed on the substrate. 
The SAW phase to frequency conversion factor as observed on the Bode plot is 
1680~Hz/$^\circ$: the resulting frequency variation is translated
into a bound mass thanks to the calibrated sensitivities -- in our
case 165~g/cm$^2$ or 145~g/cm$^2$ depending on the SAW device being used.
The resulting hydrated collagen adsorbed films are estimated to weight
1.75$\pm$0.15 and 2.10$\pm$0.20~$\mu$g/cm$^2$ for the 30 and 300~$\mu$g/ml
solutions respectively. 

\begin{figure}[h!tb]
\begin{center}
\includegraphics[width=\linewidth]{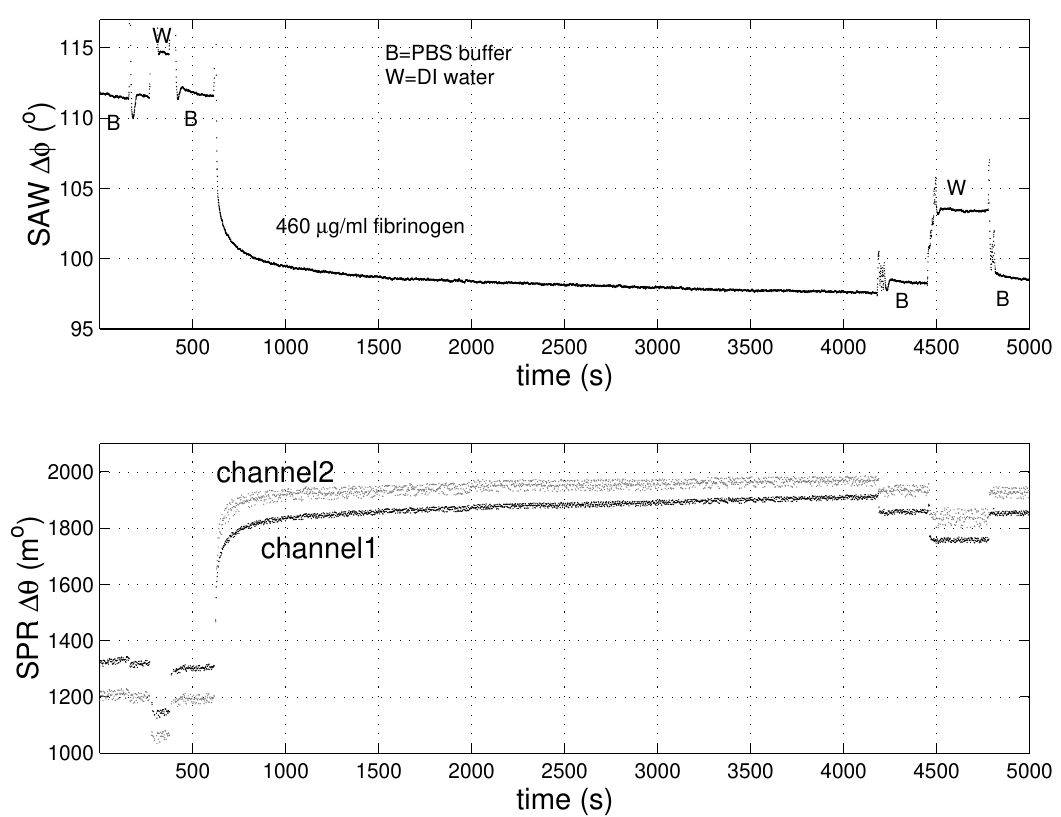} 
\end{center}
\caption{460~$\mu$g/ml fibrinogen adsorption curves. Top: phase shift monitored
at the constant frequency of 123.2~MHz, which can then be converted to an
equivalent frequency shift as would be observed in a Phase-Locked Loop oscillator
configuration via the linear phase to frequency relationship. Fibrinogen
adsorption lead to an insertion loss rise by 0.5~dB, from 23.0~dB prior
to adsorption to 23.5 after adsorption. Bottom: SPR
dip position angle shift. The sequence is PBS buffer (2 solution
exchanges), DI water, PBS, fibrinogen in the same PBS buffer, PBS rinsing 
step, DI water and PBS.}
\label{fb}
\end{figure}

Fig. \ref{fb} displays a similar measurement on 
460~$\mu$g/ml fibrinogen. A similar sequence of conditioning steps as 
previously presented for collagen were applied. PBS washing results in a
stable loss of response attributable, like for collagen, to a loss of
fibrinogen molecules physisorbed to the adsorbed film. There is again no
significant desorption of adsorbed molecules. The observed shifts resulting
from fibrinogen adsorption were 14.0$\pm$0.7~$^\circ$ and 0.65$\pm$0.1~$^\circ$
for SAW and SPR respectively. For SAW, using the same relations as was
presented for collagen, the surface density of the hydrated fibrinogen films
were estimated to be 0.75$\pm$0.1 and 1.5$\pm$0.3~$\mu$g/cm$^2$ for 46
and 460~$\mu$g/ml concentrations respectively. 

\subsection{QCM-D independent validation mesurements}

\begin{figure}[h!tb]
\includegraphics[width=\linewidth]{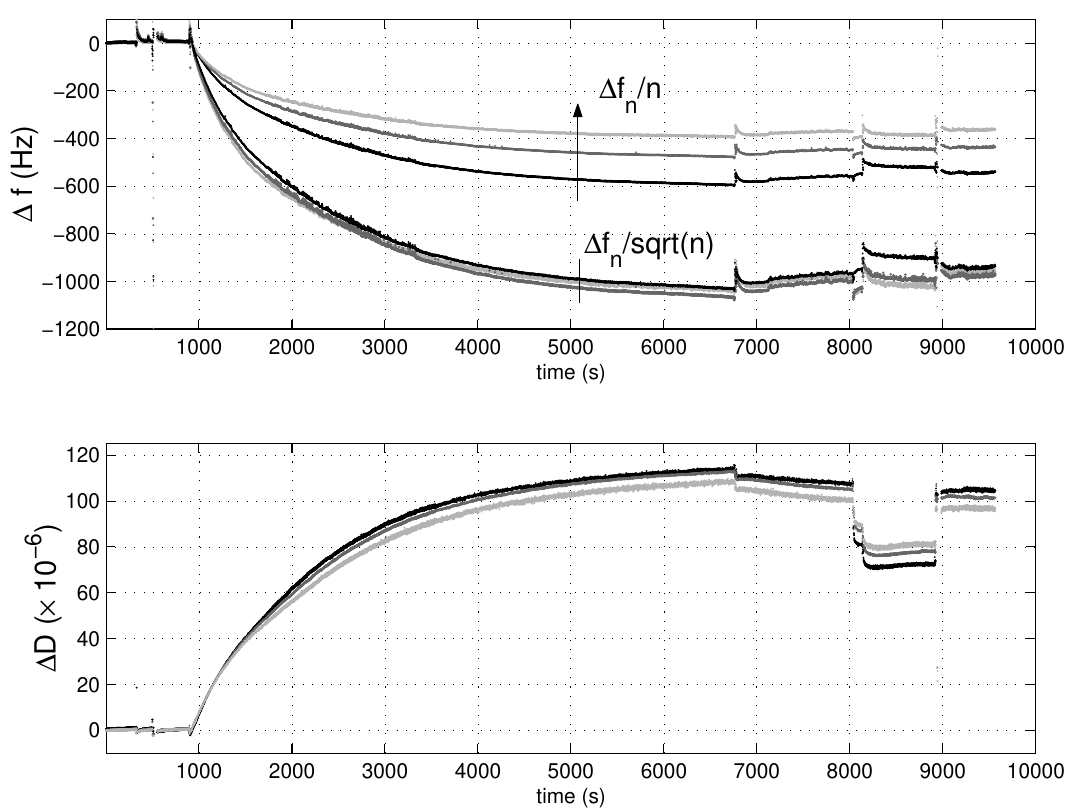}
\caption{
QCM-D 300~$\mu$g/ml collagen adsorption experiment. Top: 
normalized QCM resonance
frequency assuming a predominantly hydrodynamic interaction (scaling law as 
$1/\sqrt{n}$ with $n$ being the overtone number) and assuming a rigid layer
(scaling law as $1/n$). Notice that the latter set of curves do not overlap
and hence the rigid mass hypothesis is not correct. Bottom: QCM dissipation 
monitoring. Notice the very large damping value reached ($\Delta D > 100\times
10^{-6}$).}
\label{col_QCM}
\end{figure}

Collagen and fibrinogen adsorption were performed on methyl-terminated QCM
surfaces following the same conditions as were used for the SAW/SPR. Changes
in resonance frequency and dissipation of the first four odd overtones were
measured as a function of time. An example of a QCM-D measurement is presented 
in Fig. \ref{col_QCM} for the same 300~$\mu$g/ml collagen as
in Fig. \ref{col} (b) but on a different setup: both experiments 
(SAW/SPR and QCM) are based on a lab-made open cuvette in which 200~$\mu$l 
of the new solution is manually injected by a micro-pipette. While the 
kinetics of the SAW and SPR data can be compared, we will not attempt to 
compare it with the kinetics of the QCM since the latter was included in a
different setup -- static well geometry does influence the adsorption
kinetic \cite{faten}. 

We will however see that the asymptotic behavior
of both experiments performed separately lead to the same conclusion,
namely the large amount of water trapped in the collagen layer. Frequency
shifts are presented after normalizing by the overtone number as
classically done when assuming the adsorbed layer to be rigidly bound to the
sensing surface, and by the square root of the overtone number as would be
expected for a strong hydrodynamic interaction \cite{ECS}. The frequency
responses of the different overtones do not overlap when normalized by the
overtone number while they do when normalized by the square root of the
overtone number, indicating a predominantly viscous interaction of the QCM
with the film and surrounding solvent. Leading to the same conclusion, the dissipation rises to an
enormous value above 100$\times 10^{-6}$. Quantitatively, if this film was 
assumed to be a rigid layer, these frequency shifts would correspond to
wet masses in the range 8 to 12~$\mu$g/cm$^2$ as obtained by applying the
Sauerbrey relation. These values are far larger than the corresponding
values deduced by SAW measurements (2.10$\pm$0.20~$\mu$g/cm$^2$). This
points to the fact that SAW measurements are far less sensitive to
viscoelastic interactions with the adsorbed film than the QCM. 
All measurements for collagen and fibrinogen adsorptions at the different
concentrations were performed and the characteristic results as explained
above are summarized in Table \ref{table1}. 
The same conclusion as above can be made
in all cases, QCM measurements denoting a behavior of the
adsorbed film strongly interacting throug hydrodynamic coupling with the
bound and surrounding solvent, and the corresponding wet adsorbed masses being always
superior to those deduced by SAW. Fibrinogen shows less viscoelastic  
properties than the collagen film and lower protein concentrations result
in a less viscoelastic behavior of the adsorbed film.

\begin{table}[h!tb]
\hspace{-2cm}
{\small \begin{tabular}{|l|c|c|c||c|c|c|} \hline
Analyte (bulk           & surface & $d$ (nm)   &$x$ (\%)&$\Delta f_n/\sqrt{n}$ &$\Delta f_n/n$ & $\Delta D$ ($\times 10^{-6}$)\\
concentration, $\mu$g/ml)& density (ng/cm$^2$) & SAW/SPR  &SAW/SPR & (Hz) QCM             & (Hz) QCM      & QCM               \\ \hline
\hline
{$Cu$}                  & & $2-12$      &   ?    &    ?-1000            & NO            & {50}           \\ \hline \hline
{S-layer}               & 560$\pm$20 & $4.7\pm 0.7$&{$75\pm15$}&  NO              & 45=900        &{3-5}           \\ \hline
{CTAB}                  & 125$\pm$15 & $1.0\pm 0.1$&{100}      &  NO              & 8=160         &{0.2-0.5}       \\ \hline \hline
collagen ($30 \mu$g/ml)   & 1750$\pm$150 & $16.0\pm3.0$ & $25\pm15$ &       1000  &    NO         &  100           \\ \hline
collagen ($300 \mu$g/ml)  & 2100$\pm$200 & $19.0\pm3.0$ & $35\pm10$ &       1200  &    NO         &  $>$120        \\ \hline
fibrinogen ($46 \mu$g/ml) & 750$\pm$100  & $6.0\pm 1.5$ & $50\pm10$ & 110$\pm$5   & 55$\pm$5$\simeq 1110$ & 4-10  \\ \hline
fibrinogen ($460 \mu$g/ml)& 1500$\pm$500 & $13.0\pm2.0$ & $50\pm10$ &  NO         & 100=1700      & 8-10           \\ \hline
\end{tabular}
}
\caption{Summary of the results extracted from simultaneous SAW and SPR 
measurements, and comparison with the results obtained from QCM-D 
measurements. The ``surface density'' value is obtained by converting
the phase shift observed on the SAW device monitored in an open loop
configuration to frequency through the linear relationship monitored on
the Bode plot, and the resulting frequency shift to a mass density per
unit area by using the mass sensitivity deduced from the copper calibration.
The layer thickness $d$ and protein content of the layer $x$ are
deduced from the simulation of the SPR shift as a function of layer
thickness and optical index (assuming a protein layer optical index
of 1.45 and a protein layer density of 1.4~g/cm$^3$) and correspond
to the pair of parameters matching both the SPR and SAW data. The 
last three columns are related to QCM-D measurements and illustrate
the fact that the higher the water content in the layer under investigation,
the higher the dissipation ($\Delta D$) and the more appropriate the 
hydrodynamic model of the wave interacting with the viscous solvent
(with a 1/$\sqrt{n}$ normalization factor) over the rigid mass model (with
a 1/$n$ normalization factor). The equality sign in the 6th column
(in which the rigid mass assumption is made and in which the Sauerbrey
relationship is applied) relates the normalized frequency shift to a surface
density in ng/cm$^2$ using the proportionality factor 20.1~ng/Hz for the
4.7~MHz resonance frequency resonator.}
\label{table1}
\end{table}

Notice the excellent agreement between our results and those cited in
the literature about dry mass of collagen layers adsorbed under
similar conditions as measured by AFM and XPS: 20~nm thick layers
under water and 7-8 nm after drying \cite{cupere}, and adsorbed
amount of (dry) collagen between 0.4 and 0.8 $\mu$g/cm$^2$ are 
consistent with our 19$\pm$3~nm thick layers and a dry mass of 
$2.1\times 0.35=0.7$~$\mu$g/cm$^2$, justifying our initial assumptions
when modeling the adsorbed layer interaction with the acoustic and
optical waves.

\section{Discussion}

\subsection{Physical properties of the layers}

\begin{figure}[h!tb]
\includegraphics[width=\linewidth]{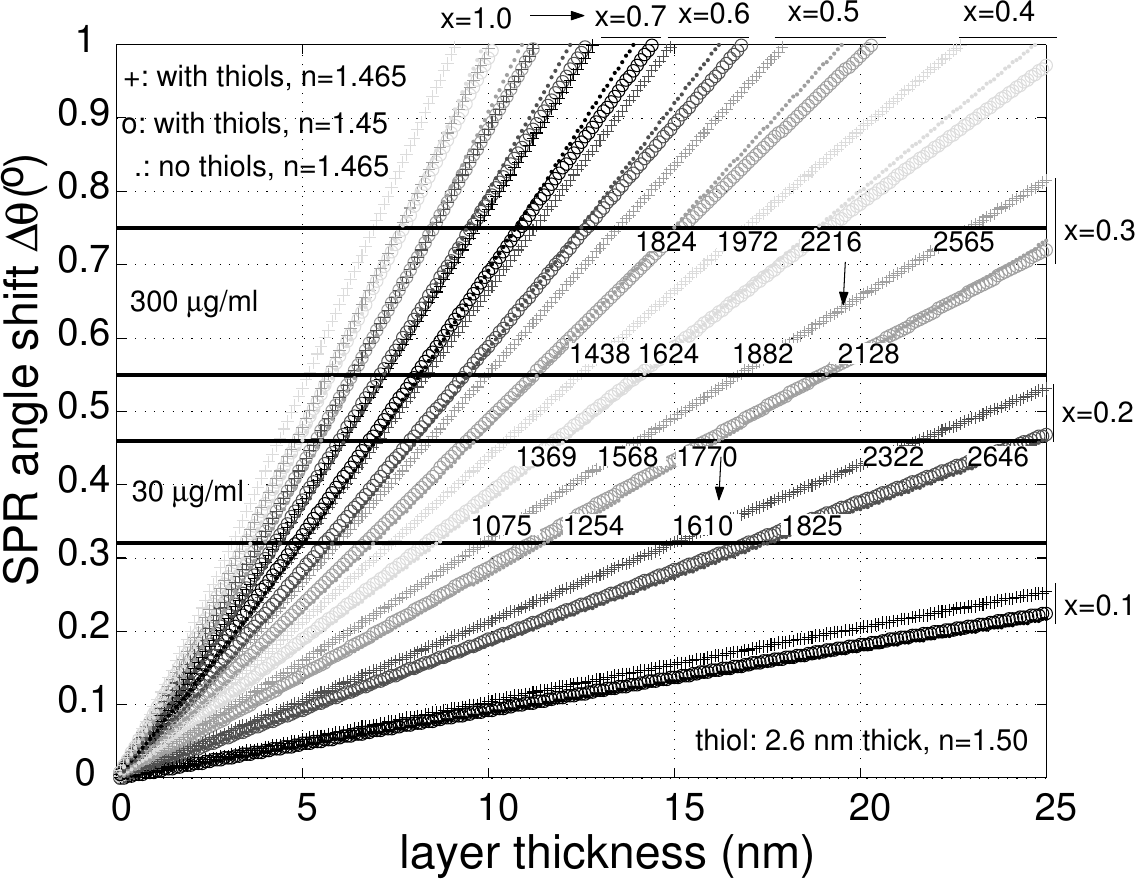}
\caption{Simulation of the angle shift (ordinate) as a function of layer 
thickness (abscissa) and protein content ratio $x$ (right and top margin)
for various layers, including or not including the thiol layer and various 
protein layer optical index (1.45 or 1.465). The mass density in ng/cm$^2$ are 
indicated within the graph for the points compatible with the observed SPR 
angle shift (390$\pm$70~m$^\circ$ for 30~$\mu$g/ml bulk collagen concentration and 
650$\pm$100~m$^\circ$ for 300~$\mu$g/ml bulk collagen concentration): the only 
point compatible with the 1750$\pm$150~ng/cm$^2$ (2100$\pm$200 respectively) 
observed by the SAW device results in a layer thickness of 16$\pm$3~nm 
(19$\pm$3~nm respectively) and a protein content ratio of 25$\pm$15\%
(35$\pm$10\% respectively). Optics simulations were performed either assuming 
a 2.6~nm thick thiol layer \cite{wim} 
(refractive index: 1.50) on the $Ti/Au$ layer followed by a variable thickness 
of collagen (mixture of protein with optical index 1.465 or 1.45 and water 
with optical index 1.33: crosses and circles), or assuming that the thiol 
layer leads to a negligible effect and not including it in the simulation, 
with a protein optical index of 1.465 (dots). The resulting mass density 
ranges (in ng/cm$^2$) are calculated for the intersection point of each of these
curves with the observed angle shift for each possible protein to water 
content ratio: all values indicated on the figure are
for solvent swollen layers, {\em including} the mass of the trapped solvent.
}
\label{simul}
\end{figure}

The observed SAW phase shift monitored at 123.2~MHz thus translates into a 
surface mass density of 1.75$\pm$0.15~$\mu$g/cm$^2$ and 
2.10$\pm$0.20~$\mu$g/cm$^2$ for bulk concentrations of collagen of 30 and 
300~$\mu$g/ml respectively. 
The SPR angle shift data are modeled following the formalism 
developed in \cite{vigoureux} in which each component of the electrical 
field is propagated through a stack a planar dielectic interfaces.
The optical indices of the metallic 
layers were taken from the literature \cite{palik} considering the
670~nm wavelength: 10~nm titanium with
$n_{Ti}=2.76+i3.84$ and 45~nm gold layer with $n_{Au}=0.14+i3.697$, 
while the optical indexes of the underlying dielectric substrate were 
taken as $n_{quartz}=1.518$ coated with 1200$\pm$100~nm silicon dioxide 
$n_{SiO_2}=1.45$. While only the last dielectric layer ($SiO_2$) below 
the metallic layer is necessary to estimate the position of the plasmon 
peak, we include the full stack of planar layers in our simulation in
order to include possible interference effects. The optical index of the
protein $n_{protein}$ is assumed to be in the 1.45 to 1.465 range, as is 
the case for most polymers
\cite{hook2,marsch,liedberg}. The properties of the protein
layer, supposed to be homogeneous, are deduced from the assumption that 
a weighted 
average of its components (water and protein) can be used: if
the ratio of protein in the layer is $x$, then the optical index
of the layer is assumed to be $n_{layer}=x\cdot n_{protein}+(1-x)\cdot n_{water}$
\cite{clusters} ($n_{water}=1.33$) and its density $\rho_{layer}=
x\cdot \rho_{protein}+(1-x)\cdot\rho_{water}$ ($\rho_{water}=1$~g/cm$^3$ and
$\rho_{protein}$ is assumed to be equal to 1.4~g/cm$^3$
range \cite{caruso,kasemo3,hook2,marsch}). The latter value is used
in our simulations. The methodology for extracting the physical parameters
of the protein layers is the following: we first simulate the predicted
SPR dip angle shift as a function of layer protein content $x$ and layer
thickness $d$, as shown in Fig. \ref{simul} where we display the SPR angle
shift due to the added dielectric layer formed by the proteins ({\it i.e.}
dip angle with solvent before adsorption subtracted from the dip angle after 
adsorption). 

We mark on such a plot
the experimentally observed angle shift, which leads to a set of possible
\{$x$, $d$\} couples compatible with the SPR experiment. For each of these
set we compute the surface density of the layer $d\times \rho_{layer}$ and 
compare it
with the value $\Delta m/A=\Delta f/(S\times f_0)$ experimentally observed 
with the SAW measurement. A single set \{$x$, $d$\} is compatible with both 
the SAW and SPR data. Notice that the resulting mass density 
$d\times \rho_{layer}$ 
{\em includes} the mass of trapped solvent in the layer.

The results of both SPR and SAW signals can only be compatible with a water 
content of the collagen layers in the range 75\%$\pm 15$\% for 30~$\mu$g/ml
concentration ($x$=0.25$\pm$0.15) and 65\%$\pm 10$\% for 300~$\mu$g/ml 
($x$=0.35$\pm$0.1), with a lower water content for higher bulk collagen 
concentrations, and correspond to the only set of parameters compatible with 
the SPR angle shift and
the mass variation observed by the SAW device, assuming a purely rigid
mass effect and negligible viscous interaction. The very high water 
content obtained here is confirmed by the large dissipation monitored 
on the QCM-D. 

As was shown earlier in \cite{ECS} and by others \cite{kasemo2},
a large dissipation shift indicates strong hydrodynamic interaction of
the vibrating QCM surface with the surrounding solvent, and is 
confirmed by the overtone scaling law $\Delta f_n/\sqrt{n}=$constant
(as opposed to the expected $\Delta f_n/n=$constant for a rigid layer).
One also sees here the strong overestimate of the bound protein mass
if assuming a rigid mass and applying the Sauerbrey proportionality
relationship between mass and frequency shift. The estimated collagen
film thickness in the 15 to 20~nm range (Fig. \ref{simul}) is in 
agreement with AFM topography maps obtained ex-situ under similar 
conditions \cite{denis}.

Table \ref{table1} summarizes our measurement results on collagen and
fibrinogen layers following the data processing methodology described
in the previous paragraph. We observe that the parameters extracted
from the SAW/SPR simultaneous measurements are in good qualitative 
agreement with the QCM-D results: a layer leading to a high dissipation
increase ($\Delta D>10\times 10^{-6}$) and an overtone normalization
law with the square root of the overtone number is characterized by 
a high water content, as seen with collagen. On the other hand a small 
dissipation increase and a normalization law of the QCM frequencies with
the overtone number are indicative of a low water content and a layer
behaving as a rigid mass bound to the surface, as observed with CTAB
and the S-layer. Fibrinogen is an intermediate case, for which the
QCM frequency does not scale well with the overtone number nor with the
square root of the overtone number, and for which the dissipation increase
is average. We indeed find a rather low water content for such layers.
The uncertainty on the
collagen layer density and hence thickness is large: as opposed to highly
ordered layers such as those formed by self-assembled monolayers of thiols
or by CTAB, we do not expect highly disorganized polymer-like films of
fibrous proteins to form exactly similar structures from one experiment
to another. The physical property (water content and hence
density and optical index) results on the other hand are consistent from 
one experiment to another, hence the need for the simultaneous measurements
at the same location by both optical and acoustic methods (and eventually
additional methods such as scanning probe microscopy \cite{fib2}) in order to 
obtain statistically significant properties of the film.

\subsection{Kinetics analysis}

\begin{figure}[h!tb]
\begin{tabular}{cc}
\includegraphics[width=0.49\linewidth]{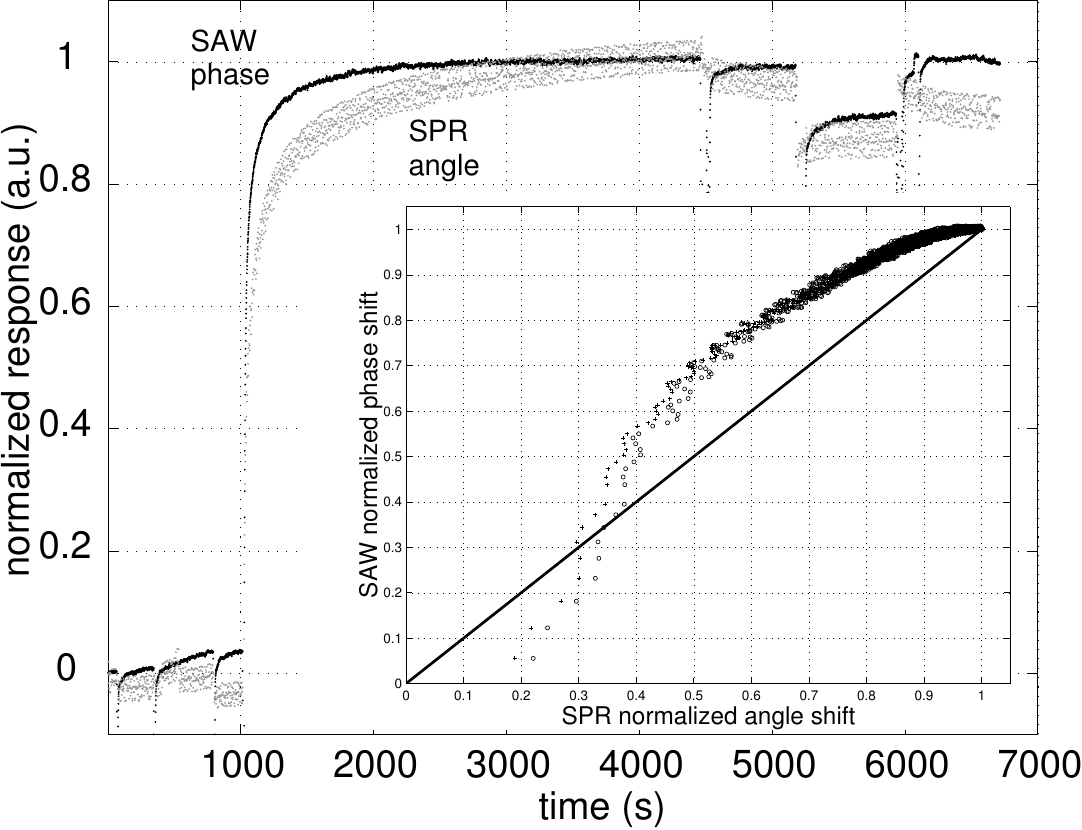} & 
\includegraphics[width=0.49\linewidth]{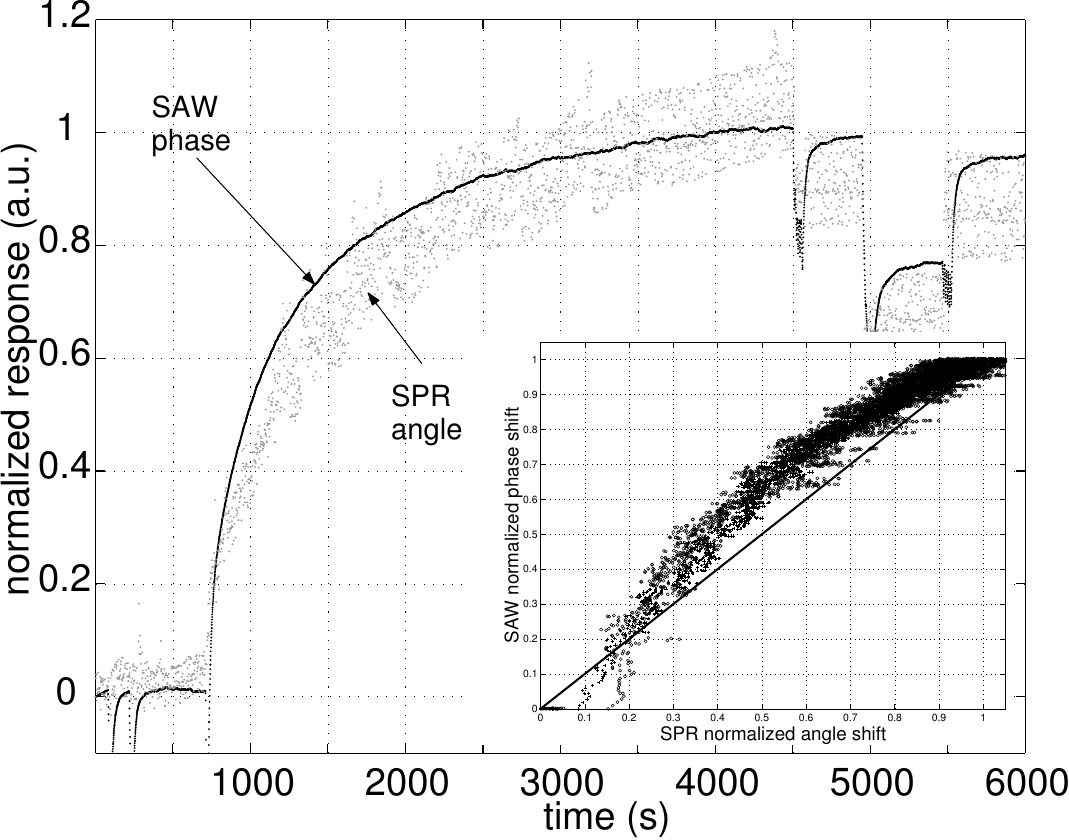} \\
(a) & (b) \\
\end{tabular}
\caption{
Comparison of the kinetics of collagen adsorption as monitored by
an acoustic (SAW) method and an optical (SPR) method, for bulk
concentrations of (a) 30~$\mu$g/ml and (b) 300~$\mu$g/ml. Notice that the
{\em same} hydrophobic-thiols coated gold area of monitored by both techniques.
The kinetics as observed by acoustic methods appears faster than that
observed by optical methods: as explained in the text, the mass of solvent
swollen layer is overestimated by acoustic methods and underestimated
by optical methods, leading to differences in the observed kinetics during
layer formation. Inset: normalized SAW response as a function of the
normalized SPR response. A rigid interaction in which both signal are
proportional to the adsorbed protein quantity would lead to all points
lying on the line of slope 1 (see text).}
\label{col_kin}
\end{figure}

We attempt now to provide some hint at differences in the kinetics as
observed by acoustic and optical methods. We compare the kinetics as
monitored by SAW and SPR by normalizing the response of the sensor
once the saturation level is reached, just after the rinsing buffer
step. We thus plot in Figs. \ref{col_kin} (a) and (b) the curves

\begin{equation}
\frac{S(t)-S(t_0)}{S(t_f)-S(t_0)}\label{eq1}
\end{equation}

where $t_0$ is the adsorption
starting time, $t_f$ the time at which a saturation signal has been
reached and the protein solution is replaced by buffer, while $S(t)$ 
is a signal which can either be in our case the SAW phase shift $\Delta\phi$
or the SPR angle shift $\Delta\theta$. We now first analyze the expected 
behavior of the normalized quantities of the two signals when assuming
a rigid mass interaction and the model presented previously here, and will
then discuss some of the possible causes of deviation from this expected
behavior.

If we first assume that the SAW signal shift is proportional to the
bound mass -- including the trapped solvent -- and that the SPR signal
is proportional to the added protein quantity, then $\Delta\phi\propto
d\times \rho(x)$ and $\Delta\theta\propto d\times n(x)$ where $d$ is 
the common layer thickness 
measured by both techniques. By introducing these proportionality
relationships in Eq. \ref{eq1}, we obtain normalized quantities expressed
as $\frac{\rho(x)-\rho_0}{\rho_f-\rho_0}$ for the SAW phase and 
$\frac{n(x)-n_0}{n_f-n_0}$ for the SPR angle, where an $f$ index indicates
the value of the quantity at the date at which saturation is reached and
index $0$ indicated the beginning of the adsorption. 

If we furthermore assume, as has been done previously in the quantitative
analysis of the data, that $\rho_{layer}=x\cdot \rho_{protein}+ (1-x)\cdot\rho_{water}$ 
and $n_{layer}=x\cdot n_{protein}+ (1-x)\cdot n_{water}$, and considering additionally
that $\rho_0=\rho_{water}$ and $n_0=n_{water}$, we finally find that both
normalized quantities express as 
$\frac{x\cdot(\rho_{protein}-\rho_{water})}{\rho_f-\rho_{water}}$ and
$\frac{x\cdot(n_{protein}-n_{water})}{n_f-n_{water}}$. Since both 
techniques, SAW and SPR, are probing the same surface, not only do they
measure a common layer thickness, but also a common protein to water ratio,
so that a common final value $x_f$ can be attributed to both $n_f$ and
$\rho_f$, and finally the two expressions simplify and give 
$\frac{x\cdot(\rho_{protein}-\rho_{water})}{x_f\cdot(\rho_{protein}-
\rho_{water})}=\frac{x}{x_f}$ for the normalized SAW data and
$\frac{x\cdot(n_{protein}-n_{water})}{x_f\cdot(n_{protein}-n_{water})}=
\frac{x}{x_f}$.

\begin{figure}[h!tb]
\includegraphics[width=\linewidth]{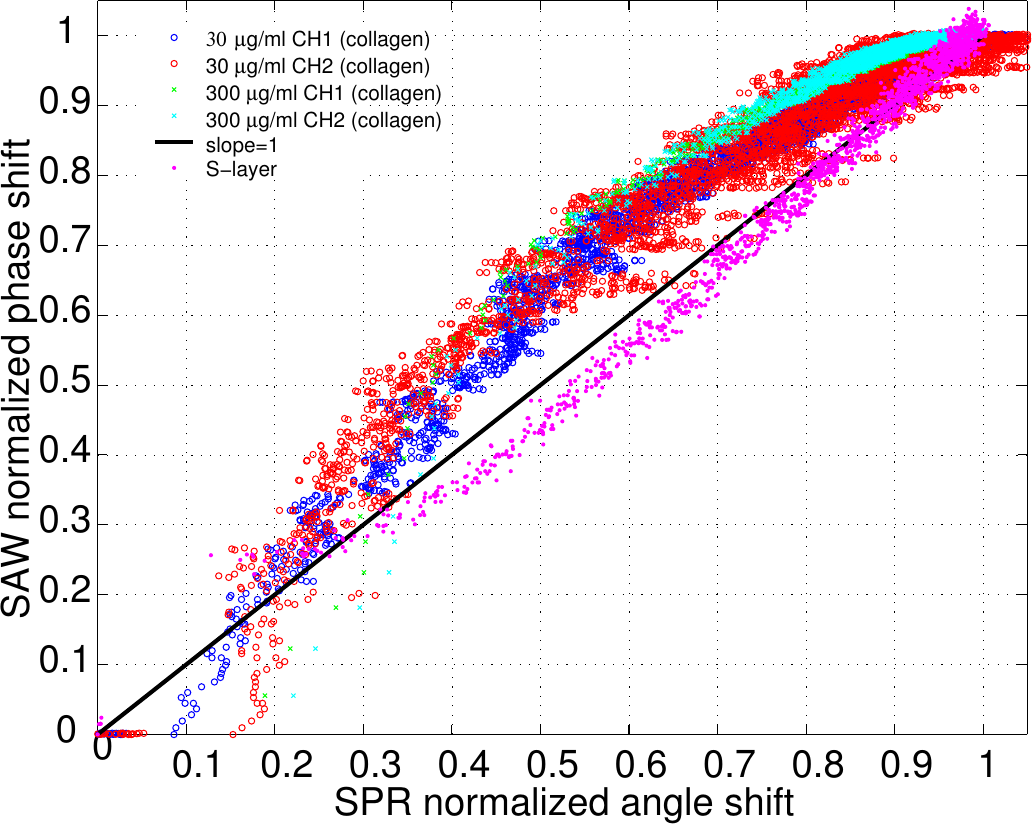}
\caption{Normalized signal from acoustic (SAW) and optical (SPR) measurements:
if both techniques were measuring the same adsorbed mass property, the data
would be lying on the $y=x$ line, as seen for the rigid S-layer. For collagen
and fibrinogen, all datasets lie above $y=x$.}
\label{sumyx}
\end{figure}

In conclusion of this calculation, we see that following two assumptions
about the proportionality of the SAW and SPR signals with the physical 
properties of the layer and the evolution of the physical property of the
protein layer as a function of water content, we reach the same law for
the evolution of the normalized SAW and SPR signals. We test this
result by eliminating the time parameter and plotting the SAW normalized phase 
shift as a function of SPR normalized angle shift. We predict that all
points should lie on the line of slope 1. The S-layer data shown previously
\cite{JAP} do not show 
any significant difference in the kinetics as detected by the optical and 
acoustic methods, and hence indeed all normalized points fall close to this
line (Fig. \ref{sumyx}). On the other hand the solvent swollen fibrinogen and collagen adsorbed layers
presented here seem to display a faster kinetics as detected by the acoustic 
method than the one detected by the optical method. Therefore we see in inset of
Fig. \ref{col_kin} (a) and (b) that all points lie above the line of
slope 1. Such a graph provides a quantitative relationship between 
measurements obtained by acoustic and optical methods.

From the calculation we just presented, we can attribute the discrepancy
of the normalized point distribution to either of the two assumptions: either
the SAW signal is not strictly proportional to the bound mass (assumption
$\Delta\phi\propto d\times \rho_{layer}$) or the simple linear law relating
the physical properties of the layer (density and optical index) to the
physical properties of the individual components of the layer -- protein
and water -- is not accurate. In view of the previously published
literature \cite{kasemo3,viscous_gizeli}, we favor the first possibility 
and propose a possible contribution of the hydrodynamic interaction of the 
collagen layer to the SAW signal. The acoustic signal overestimates the bound
mass by being affected by the bound solvent as well, while the optical
method under-estimates the bound mass due to the lowered optical index of
the adsorbed layer associated with the bound solvent.

Throughout this analysis, neither the viscoelasticity of the adsorbed layer
or of the surrounding solvent, nor the experimentally measured acoustic losses,
have been considered. The shortcoming of these simplifications might be solved
by adding a set of charts relating the frequency shift with layer thickness
and viscosity, and most significantly insertion loss with these
parameters. Such an investigation was endeavoured in \cite{lamia,lamia2},
in the former reference for solvent only (no adsorbed layer) and in the latter
for a single layer in air, since the coupled phase/magnitude shift induced by both
mass loading and viscoelastic coupling makes the separation of each
contribution complex when considering finite thickness layers coated by a viscous 
solvent. The insertion loss evolution observed between the beginning and end of each 
experiment has been indicated in the caption of each experimental SAW chart result 
for the reader to assess the validity of this assumption: while a minute insertion loss
change in the case of fibrinogen (0.5~dB) hints at the validity of our assumptions,
the large (3.3--6.5~dB) insertion loss observed during collagen adsorption hints at
the need for a more complex analysis to grasp all the subtelties of the adsorption
process.

\section{Conclusion}

We have shown how a two steps measurement, first using copper
electrodeposition for instrument calibration followed by a measurement on
the actual biological layer under investigation, on a combined SAW/SPR
instrument leads to an estimate of the water content and layer
thickness. While each individual technique cannot achieve such
identification, both common parameters (water content ratio and layer
thickness) can be identified with the combination. Such results are
especially useful for protein layers with high water content for which
the hydrodynamic interaction can become predominant over rigid adsorbed mass.
We have verified the calibration process using CTAB, confirming the
120~ng/cm$^2$ surface density due solely to the densely organized monolayer,
and illustrated the technique with an estimate of the water content
of collagen, leading to a 70\%-water/30\%-protein result, which is in
agreement with the large dissipation increase (quality factor decrease) 
observed by QCM-D. Furthermore, QCM-D overtone normalization factor with
such solvent swollen layers is the inverse of the square root of the 
overtone number (and not the inverse of the overtone number as would be 
the case for a rigid layer). Applying the proportionality factor between
QCM frequency shift and surface mass density as predicted by the Sauerbrey
relationship, assuming a rigidly bound mass, leads to a strong overestimate of 
the adsorbed mass. We have also shown that fibrinogen layers have a water 
content of 50$\pm$10~\%, compatible with a lower dissipation factor increase 
in the QCM-D measurements. Finally, we have discussed the comparison of 
kinetics as observed by optical and acoustic techniques and propose that
the discrepancies observed in the case of solvent swollen layers might be
attributed to minor hydrodynamic interactions of the SAW with the protein layer.
Introducing in this analysis the SAW insertion losses would allow to resolve
one more unknown, namely the adsorbed layer dynamic viscosity.

\section{Acknowledgments}

All literature references were fetched on the Library Genesis at \url{gen.lib.rus.ec},
whose service is invaluable to our research activities. Funding was provided
by European Comission Framework Program grants FP5-IST PAMELA and 
FP7-ICT LoveFood.



\section*{References}

\bibliography{biblio.bib,biblio2016/biblio2016.bib}

\end{document}